\documentclass[aps, prl, 10pt, twocolumn, superscriptaddress,floatfix]{revtex4-1}
\usepackage{bbold}
\usepackage{graphicx}
\usepackage{ulem}
\usepackage{amsmath,amssymb,amsfonts}
\usepackage[percent]{overpic}
\usepackage{braket}
\usepackage{bm}
\usepackage[breaklinks=true,colorlinks,citecolor=blue,linkcolor=blue,urlcolor=blue]{hyperref}
\usepackage[usenames, dvipsnames]{color}
\usepackage{mathrsfs}
\usepackage{ulem}
\usepackage{xcolor}

\begin{document}
\title{Genuine quantum advantage in non-linear bosonic quantum batteries}
\author{Gian Marcello Andolina}
\altaffiliation{These two authors contributed equally.\\
\href{mailto:gian-marcello.andolina@college-de-france.fr}{gian-marcello.andolina@college-de-france.fr}\\
\href{mailto:vittoria.stanzione@phd.unipi.it}{vittoria.stanzione@phd.unipi.it}}
\affiliation{JEIP, UAR 3573 CNRS, Coll\`ege de France, PSL Research University,
11 Place Marcelin Berthelot, F-75321 Paris, France}
\author{Vittoria Stanzione}
\altaffiliation{These two authors contributed equally.\\
\href{mailto:gian-marcello.andolina@college-de-france.fr}{gian-marcello.andolina@college-de-france.fr}\\
\href{mailto:vittoria.stanzione@phd.unipi.it}{vittoria.stanzione@phd.unipi.it}}
\affiliation{Dipartimento di Fisica dell'Universit\`a di Pisa, Largo Bruno Pontecorvo 3, I-56127 Pisa,~Italy}
\author{Vittorio Giovannetti}
\affiliation{NEST, Scuola Normale Superiore and Istituto Nanoscienze-CNR, I-56126 Pisa, Italy}
\author{Marco Polini}
\affiliation{Dipartimento di Fisica dell'Universit\`a di Pisa, Largo Bruno Pontecorvo 3, I-56127 Pisa,~Italy}
\affiliation{ICFO-Institut de Ci\`{e}ncies Fot\`{o}niques, The Barcelona Institute of Science and Technology, Av. Carl Friedrich Gauss 3, 08860 Castelldefels (Barcelona),~Spain}
\begin{abstract}
Finding a quantum battery model that displays a genuine quantum advantage, while being prone to experimental fabrication, is an extremely challenging task. In this Letter we propose a deceptively simple quantum battery model that displays a genuine quantum advantage, saturating the quantum speed limit. It consists of two harmonic oscillators (the charger and the battery), coupled during the non-equilibrium charging dynamics by a {\it non-linear} interaction. We first present the model, then certify the genuine quantum advantage, and finally briefly discuss how the battery can be fabricated through the use of superconducting circuits. 
\end{abstract}
\date{\today}
\maketitle

\noindent {\color{blue}{\it Introduction}}.---Quantum thermodynamics~\cite{Kosloff_Entropy_2013,Vinjanampathy_ContempPhys_2016,Goold_JPA_2016,Lostaglio_RepProgPhys_2019,Binder_Book_2018,Deffner_Book_2019,Bhattacharjee_EPJB_2021,Myers_AVS_2022,Arrachea_RepProgPhys_2023} is an active research field where genuine quantum effects 
are sought in a variety of meso- and nano-devices including heat engines and refrigerators~\cite{Giazotto_RMP,Pekola_RMP}. Seeking quantum effects in thermodynamic processes is a far from trivial task. 
As Enrico Fermi clearly explained in his lectures held at Columbia University (NY) during the summer session of 1936~\cite{Fermi1956}, 
``in pure thermodynamics, the fundamental laws are assumed as postulates based on
experimental evidence, and conclusions are drawn from them without entering into the kinetic mechanisms of the phenomenon''.
Thermodynamics has therefore a universal character, offering predictions that are valid for both classical and quantum settings.
In order to find a {\it genuine quantum advantage} (GQA) in the context of thermodynamics, one clearly needs to transcend equilibrium conditions and study the {\it non-equilibrium} dynamics of quantum systems.
In this context, quantum batteries, first introduced in 2013 by Alicki and Fannes~\cite{Alicki13}, have recently attracted a great deal of attention~\cite{Binder15, Campaioli17,Campaioli_RMP_2024}.

In general terms, a closed quantum battery is a quantum mechanical system with a discrete energy spectrum of finite bandwidth, which can be charged---i.e.~prepared in an excited energy state $\rho$ such that ${\rm Tr}[{\cal H}_{\rm B}\rho]$ is larger than its ground-state energy---via unitary operations that may temporarily generate coherences between its eigenstates. Generalizations to open quantum batteries are of course possible~\cite{Campaioli_RMP_2024} but not the subject of this work. Other operations, such as work extraction, can also be carried out on a quantum battery via unitary controls, but in this Letter we will solely focus on the non-equilibrium charging dynamics.

As classical batteries, also quantum batteries are bound to have a maximum capacity (i.e.~a maximum of the stored energy)  and a maximum charging power. An extremely useful bound on power was derived by Juli\`a-Farr\'e et al.~\cite{JuliaFarre_PRR_2020} through a quantum geometric approach. Bounds on the charging dynamics of a quantum battery can also be casted by using the so-called ``quantum speed limits''~\cite{M-L, M-T, Defner17, Giovannetti2003a}. If a quantum battery saturates the previous bounds, it is said to display a GQA~\cite{QAdv}.

Aim of this Letter is to investigate the non-equilibrium charging dynamics of many-body quantum batteries,  i.e.~quantum batteries based on interacting quantum many-body systems. The key point is that interactions among the battery elements can in principle generate quantum correlations during the non-equilibrium dynamics, leading to a GQA. With this target in mind, several quantum battery models have been proposed in the past few years. Examples include Dicke quantum batteries~\cite{Ferraro_PRL_2018,Andolina18,QvsC,Yang_PRB_2024} and quantum batteries based on one-dimensional Heisenberg spin chains~\cite{Le_PRA_2018}, disordered quantum Ising chains~\cite{Rossini_PRB_2019}, and the Su-Schrieffer-Heeger model~\cite{Zhao_PRR_2022}. As discussed at length in Refs.~\cite{JuliaFarre_PRR_2020,Campaioli_RMP_2024}, none of these models, however, displays a GQA. To the best of our knowledge, indeed, only {\it one} quantum many-body battery model displaying a GQA has been discovered so far. This is the Sachdev-Ye-Kitaev (SYK) quantum battery, which was proposed in Refs.~\cite{Rossini_PRL_2020, Rosa_JHEP_2020}. 
Proposals to realize the SYK Hamiltonian~\cite{Sachdev93,Kitaev15,gu_arxiv_2019,rosenhaus_jpa_2019,Chowdhury_RMP_2022} in the laboratory have been put forward and rely on ultra-cold atoms~\cite{Danshita17} and solid-state systems~\cite{Franz_NatRevMater_2018}, such as topological superconductors~\cite{Chew17,Pikulin17} and graphene quantum dots with irregular boundaries in strong applied magnetic fields~\cite{Chen18,Kruchkov_PRB_2020,Brzezinska_PRL_2023}. Experimental evidence for the achievement of a regime of strong correlations in the latter systems has been recently reported by Anderson et al.~\cite{Anderson_PRL_2024}.  The authors of this work presented data for the thermoelectric power of these quantum dots exhibiting strong departures from the Mott formula in high magnetic fields (on the order of $10~{\rm T}$) and elevated temperatures ($T \gtrsim 10~{\rm K}$). These data are compatible with the emergence of a non-Fermi-liquid regime described by the SYK model~\cite{Shackleton_PRB_2024}. Although this is clearly a milestone result, fabricating quantum batteries from arrays of graphene quantum dots with disordered edges is still a long shot.

In this Letter, we propose a deceptively simple quantum battery model where fast charging and a GQA occur via an interaction Hamiltonian that has not been yet explored in the literature. Indeed, as we discuss at length below, we achieve a GQA by coupling two harmonic oscillators via a non-linear charging Hamiltonian. Similarly to the Dicke battery~\cite{Ferraro_PRL_2018}, also this quantum battery model can be fabricated in the laboratory by using e.g.~superconducting circuits~\cite{Blais}.

\noindent {\color{blue}{\it Model and charging protocol}}.---We begin by briefly reminding the reader about charger-based protocols in the non-equilibrium dynamics of quantum batteries. Consider two quantum systems~\cite{Andolina18}, the ``charger''  ${\rm A}$  and the ``proper battery'' ${\rm B}$, described by the Hamiltonians 
$\mathcal{H}_{\rm A}$ and $\mathcal{H}_{\rm B}$, respectively. At time $t=0$ we assume that the system is in a pure factorized state
 $|\psi(0)\rangle  \equiv | \psi\rangle_{\rm A}|0\rangle_{\rm B}$, where $|0\rangle_{\rm B}$  is
 the (zero-energy) ground state of $\mathcal{H}_{\rm B}$, and $| \psi\rangle_{\rm A}$ is taken to have an initial energy $E_{\rm A}(0) >0$. Aim of the charging protocol is to transfer energy from the charger to the battery. In order to do so, at time $t =0^+$ we switch on an interaction Hamiltonian $\mathcal{H}_{\rm int}$ between
 the two systems, keeping it on for a finite charging time $\tau$. The complete Hamiltonian is therefore $\mathcal{H}(t)=\mathcal{H}_{\rm A}+\mathcal{H}_{\rm B}+\lambda(t)\mathcal{H}_{\rm int}$, where  $\lambda(t)=1$ for $t\in[0,\tau]$ and zero elsewhere is a classical parameter that represents the external
control we exert on the system. We take $[\mathcal{H}_{\rm int}, \mathcal{H}_{\rm A} + \mathcal{H}_{\rm B}]=0$ in order to avoid injecting into the system more energy than that initially contained in the charger~\cite{Strasberg16,Andolina18}.

The energy $E_{\rm B}(\tau)$ stored in the battery at time $\tau$ and the average charging power $P_{\rm B}(\tau)$ are given by
\begin{align}
&E_{\rm B}(\tau) \equiv   \langle \psi(\tau) |\mathcal{H}_{\rm B} |\psi(\tau)\rangle~,\label{stored energy} \\ 
&P_{\rm B}(\tau) \equiv E_{\rm B}(\tau)/\tau~, \label{storing power}
\end{align}
where $|\psi(\tau)\rangle$ is the state of the system at time $\tau$.
We denote by the symbol ${\bar \tau}$ the {\it shortest} time at which the battery energy reaches its maximum value, i.e.~$E_{\rm B}(\tau) \leq E_{\rm B}(\bar{\tau})$. 

We are now ready to introduce the model we propose in this work, which reads as following ($\hbar = 1$ throughout this paper):
\begin{align}\label{H_NL}
&\mathcal{H}_{\rm A} = n \omega_0 a^\dagger a~, \nonumber \\
&\mathcal{H}_{\rm B} = \omega_0 b^\dagger b~,  \nonumber \\ 
&\mathcal{H}_{\rm int} \equiv \mathcal{H}_{\rm n} = g_n\big[a^\dagger b^n+a (b^\dagger)^n\big]~.
\end{align}
Here, $\omega_0$ is a natural frequency scale, $g_n$ is a coupling constant (with units of energy), and $n$ represents the order of the non-linear coupling between the harmonic oscillator $\mathcal{H}_{\rm A}$ describing the charger and the one describing the proper battery $\mathcal{H}_{\rm B}$. We take the initial energy of the charger to be given by $E_{\rm A}(0) = N\omega_0$, where $N$ is an integer.

\noindent  {\color{blue}{\it Charging dynamics and the quantum speed limit.}}---Consider for a moment the case $n = 1$, which represents a linear coupling between charger and battery. This model has already been studied in Ref.~\cite{Andolina18}. Choosing as initial state of the charger a Fock state with $N$ excitations, i.e.~$| \psi\rangle_{\rm A}=| N \rangle_{\rm A}$, one can easily calculated the energy stored in the battery and the average power, finding~\cite{Andolina18}:
\begin{eqnarray}
\label{stored energyoo}
E^{( 1)}_{\rm B}(\tau)&=&N\omega_0 \sin^2(g_1\tau)~, \\ 
\label{powerstored1}
P_{\rm B}^{( 1)}(\tau)&=&N\omega_0\frac{\sin^2 (g_1\tau)}{\tau}~.
\end{eqnarray}
(These results are in fact valid, irrespectively of the details of the initial state.)
The optimal charging time is $\bar{\tau}_1=\pi/(2g_1)$, which does depend on $N$. The linear $n=1$ quantum battery model therefore does {\it not} display any speed-up in the $N\gg 1$ limit.  

From now on, we therefore focus on the {\it non-linear} $n >1$ case, taking in particular $n=N$. In this case, we choose as initial state of the charger a Fock state with {\it one} excitation, i.e.~$| \psi\rangle_{\rm A}=| 1 \rangle_{\rm A}$. (Other initial states will be discussed below.) The interaction Hamiltonian in Eq.~\eqref{H_NL} couples the initial state $| 1 \rangle_{\rm A}| 0 \rangle_{\rm B}$ with the fully charged battery state,  $| 0 \rangle_{\rm A}| N \rangle_{\rm B}$. The interaction Hamiltonian therefore reads as following
\begin{equation}
\label{H_NL1}
\mathcal{H}_{N}= g_N\sqrt{N!}\left(| 1 \rangle_{\rm A}| 0 \rangle_{\rm B} \langle 0|_{\rm A} \langle N|_{\rm B}+{\rm H.c.}\right)+\Delta\mathcal{H}_{\rm N}~,
\end{equation}
where $\Delta\mathcal{H}_{\rm N}$ is an operator that annihilates the states $| 1 \rangle_{\rm A}| 0 \rangle_{\rm B}$ and $| 0\rangle_{\rm A}| N \rangle_{\rm B}$, i.e.~$\Delta\mathcal{H}_{\rm N}| 1 \rangle_{\rm A}| 0 \rangle_{\rm B}=\Delta\mathcal{H}_{\rm N}| 0 \rangle_{\rm A}| N \rangle_{\rm B}=0$. As a consequence, it is easy to calculate exactly the state of the system at time $t$. We find
\begin{align}\label{evolvedstate}
 |\psi(t)\rangle_{\rm AB} & = e^{-i N \omega_0 t}[\cos{(g_N\sqrt{N!}t)}|1\rangle_{\rm A} |0 \rangle_{\rm B}  + \nonumber \\
& i\sin{(g_N\sqrt{N!}t)}  |0 \rangle_{\rm A} |N \rangle_{\rm B}]~. 
\end{align} 
Using this result in Eqs.~(\ref{stored energyoo})-(\ref{powerstored1}), we obtain the following results:
\begin{align}\label{stored energyNL}
E_{\rm B}(\tau) & = N\omega_0 \sin^2(g_N\sqrt{N!}\tau)~, \\ \label{powerstored2} 
P_{\rm B}(\tau) & = N\omega_0 \frac{\sin^2 (g_N\sqrt{N!}\tau)}{\tau}~. 
 \end{align} 
The optimal charging time is therefore $\bar{\tau}_N=\pi/(2g_N\sqrt{N!})$. No microscopic relation exists so far between the linear coupling $g_1$ and the non-linear coupling  $g_N$. We therefore cannot fairly compare the non-linear model with the linear one, investigating whether the former displays a GQA. 

In order to overcome this obstacle and establish a fair comparison, we use the Mandelstam-Tamm quantum speed limit (QSL)~\cite{M-T}, which states that the time $\tau_{\rm QSL}$ needed for a system described by a Hamiltonian ${\cal H}$ to go from a state $|\psi(0)\rangle$ to an orthogonal state is given by
\begin{equation}\label{eq:QSL_theory}
\tau_{\rm QSL} \equiv \frac{\pi}{2 \langle\delta {\cal H}\rangle}_{\psi(0)}~.  \end{equation}
Here, we have introduced the variance $\langle\delta {\cal H}\rangle_{{\psi}}$ of $\mathcal{H}$ on a state $\ket{\psi}$ as $\langle\delta {\cal H}\rangle_{{\psi}} \equiv \sqrt{\langle \psi |\mathcal{H}^2|\psi\rangle - \langle \psi|\mathcal{H}|\psi\rangle^2 }$. (Below, in obtaining Eq.~\eqref{eq:QSL}, we have evaluated the variance over the initial state, ${{\ket{\psi(0)}}=| 1\rangle_{\rm A}| 0 \rangle_{\rm B}}$.) In what follows, we impose~\cite{Campaioli17,Binder15}  that both models, the linear and the non-linear one, have the same $\tau_{\rm QSL}$ and thus the same variance $\langle\delta {\cal H}\rangle_{\psi(0)}$. Imposing that $\tau_{\rm QSL}$ is the same in both cases protects us from potentially spurious effects that may accelerate the dynamics of the non-linear model. For the linear case, and with the aim of evaluating $\tau_{\rm QSL}$, we set $n=1$ in Eq.~(\ref{H_NL}) and use ${\cal H} \to {\cal H}_{\rm L} \equiv {\cal H}_{\rm A} + {\cal H}_{\rm B} +{\cal H}_1$ in Eq.~(\ref{eq:QSL_theory}). Similarly, for the non-linear case, we set the order of non-linearity at the value $n=N$ and use ${\cal H} \to {\cal H}_{\rm NL} \equiv {\cal H}_{\rm A} + {\cal H}_{\rm B} +{\cal H}_N$ in Eq.~(\ref{eq:QSL_theory}). Evaluating the two variances  we find $\langle\delta {\cal H}_{\rm L}\rangle^2_{\psi(0)} = N g_1^2$ and $\langle\delta {\cal H}_{\rm NL}\rangle^2_{\psi(0)}  =N! g_N^2$. Imposing their equivalence yields the desired relation between $g_1$ and $g_N$, i.e.
\begin{equation}\label{eq:mapping}
g_N= \frac{g_1}{\sqrt{(N-1)!}}~,
\end{equation}
and the QSL time
\begin{equation}\label{eq:QSL}
\tau_{\rm QSL} = \frac{\pi}{2 \sqrt{N} g_1}~.
\end{equation}

Replacing the crucial result (\ref{eq:mapping}) in Eqs.~(\ref{stored energyNL})-(\ref{powerstored2}), we find the following figures of merit:
\begin{align}\label{stored energyNL1}
E_{\rm B}(\tau) &= N\omega_0 \sin^2(g_1\sqrt{N}\tau)~, \\ 
\label{powerstoredNL}
P_{\rm B}(\tau) &= N\omega_0\frac{ \sin^2 (g_1\sqrt{N}\tau)}{\tau}~.
 \end{align} 
Finding the shortest time $\bar{\tau}$ at which $E_{\rm B}(\tau)$ is maximal (which is the optimal charging time mentioned above) yields
\begin{equation}\label{time}
\bar{\tau} = \frac{\pi}{2 \sqrt{N} g_1}~.
\end{equation}
We clearly see that $\bar{\tau}$ becomes {\it shorter} as $N$ increases, signalling a speed-up of the charging dynamics in the limit $N\gg 1 $. We also note that the result in Eq.~(\ref{time}) {\it coincides} with the QSL time reported above in Eq.~\eqref{eq:QSL}. Since our non-linear charging model saturates the QSL bound, i.e.~since $\bar{\tau} = \tau_{\rm QSL}$, we have a preliminary strong indication that the asymptotic speed-up displayed by Eq.~(\ref{time}) for $N\gg 1$ has a genuine quantum origin. We will return to this point below, showing with two more methods that the model (\ref{H_NL}) displays a GQA.

The advantage in the non-linear battery model stems from the fact that the first orthogonal state, to which the initial state is rotated, is exactly the one in which energy is transferred from the charger into the battery. (In fact, this is the only state of the system that is orthogonal to the initial state, which can be reached through time evolution since this conserves the total number of excitations.) In contrast, in the linear model, the first orthogonal configuration (and thus the one reachable within the time set by the QSL) is the one where one excitation is taken away from the charger and transferred into the battery (clearly this does not provide the battery with the maximum amount of energy).

Substituting the optimal charging time~\eqref{time} into Eq.~\eqref{stored energyNL1} and~\eqref{powerstoredNL}, we find
\begin{align}
\label{energy}
E_{\rm B}(\bar{\tau}) &=N\omega_0~,\\ 
\label{power}
P_{\rm B}(\bar{\tau}) &= \frac{2}{\pi} \omega_0 g_1 N^{3/2}= \frac{E_{\rm B}(\bar{\tau})}{\bar{\tau}}~.
\end{align}
We notice that, not only the optimal charging time saturates the QSL, but also all the energy is completely transferred from the charger to the battery. Furthermore, the power shows a super-linear scaling with $N$, stemming from a GQA in the optimal charging time $\bar{\tau}\propto N^{-1/2}$. 

\begin{figure}[t]
\centering
\vspace{1.em}
\begin{overpic}[width=1 \columnwidth]{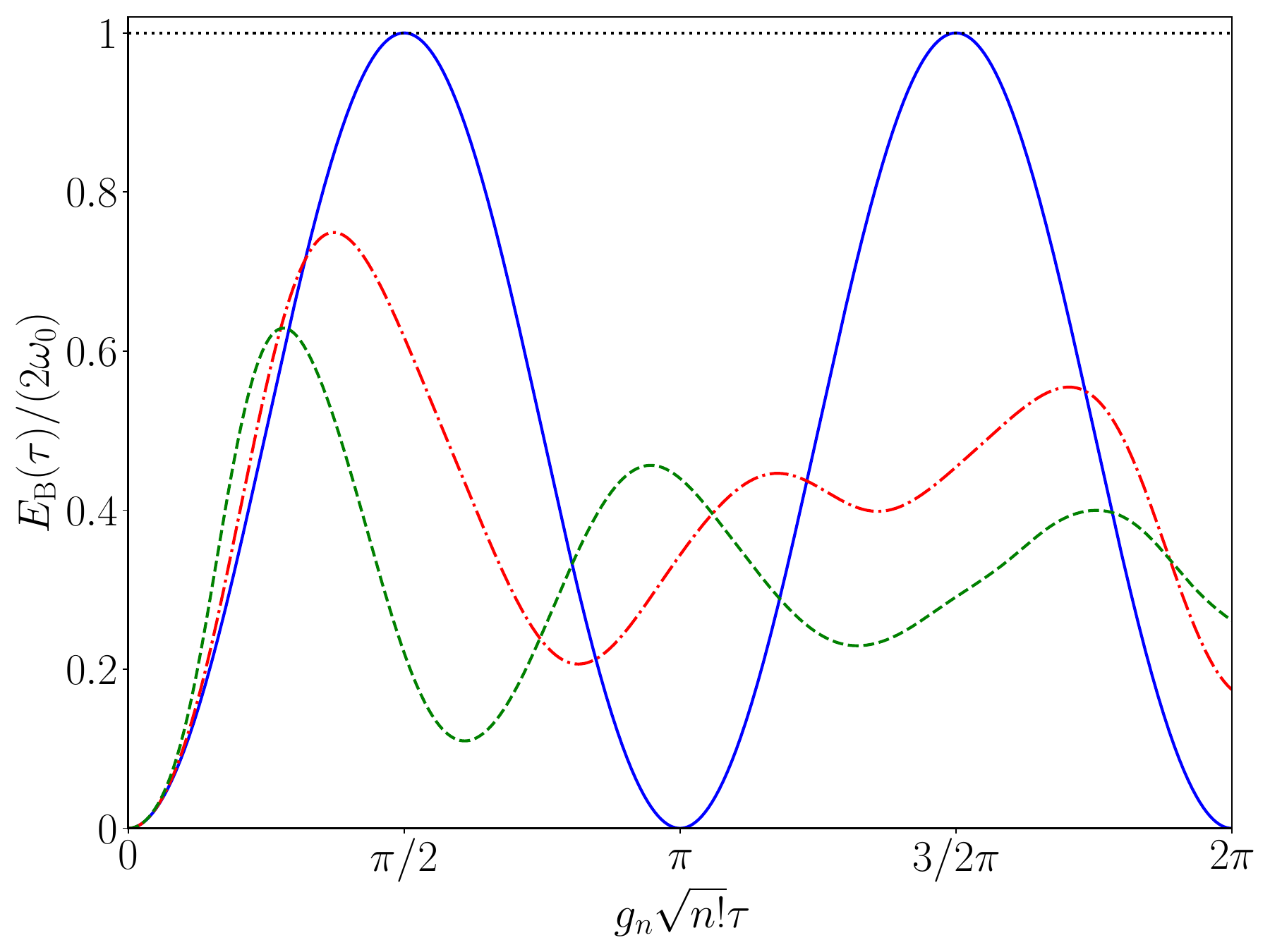}\put(6,73){ }\end{overpic}
\caption{(Color online)  Energy  $E_{\rm B}(\tau)$ (in units of $2\omega_0$) stored in the battery as a function of time $\tau$, in units of $1/(g_n \sqrt{n!})$. Data in this figure have been obtained by setting the degree $n$ of non-linearity to the value $n = 2$. Different curves refer to results obtained for three different choices of the initial state of the charger: Fock state (blue solid line), coherent state (red dash-dotted line), and squeezed vacuum state (green dashed line). The three state are fixed to have the same initial energy, $E_{\rm A}(0)=2\omega_0$. The horizontal (dotted black) line represents the perfect energy transfer case, i.e.~$E_{\rm B}(\tau)=2\omega_0$.
\label{fig:energycomparison}}
\end{figure}

We now discuss the dependence of results (\ref{time}), (\ref{energy}), and~(\ref{power}) on the initial state. If one starts from an initial state of the charger which is not a Fock state ($| \psi\rangle_{\rm A} \neq | 1 \rangle_{\rm A}$), the maximum energy in the battery will not be the maximum one expressed in Eq.~\eqref{energy}. In this sense, the Fock state is optimal in that it is the only state that is able to fully charge the battery, as shown in Fig.~\ref{fig:energycomparison}. We also emphasize that, once the variances $\langle\delta {\cal H}_{\rm L}\rangle^2_{\psi(0)}$ and $\langle\delta {\cal H}_{\rm NL}\rangle^2_{\psi(0)}$ are imposed to be equal, the Fock state is also optimal from the point of view of the charging time, given that it saturates the speed limit.  

Suppose now that the initial state of the charger is a generic state with a given number $M$ of excitations. The initial state of the battery will always be taken to be the ground state $|0\rangle_{\rm B}$.
Since $[\mathcal{H}_{\rm int}, \mathcal{H}_{\rm A} + \mathcal{H}_{\rm B}]=0$, $\mathcal{H}_{\rm int}$ can only exchange excitations between A and B. Hence, the initial state of the system evolves according to $U_t\ket{M}_{\rm A}\ket{0}_{\rm B}=\sum_{m=0}^M c_{m,M}(t)  \ket{M-m}_{\rm A}\ket{N m}_{\rm B}$, where $U_t=e^{-i\mathcal{H}t}$ is the unitary evolution generated by the full Hamiltonian $\mathcal{H}=\mathcal{H}_{\rm A}+\mathcal{H}_{\rm B}+\mathcal{H}_{\rm int}$ and  $c_{m,M}(t) $ are unknown function of time.
The density matrix of a generic mixed state (being the battery empty of energy) can be written as: $\rho=\sum_{M=0}^\infty \sum_{K=0}^{\infty}\rho_{M,K}\ket{M}_{\rm A}\ket{0}_{\rm B}\bra{0}_{\rm B}\bra{K}_{\rm A}$. Thus, the energy stored energy at time $t$ is
\begin{equation}\label{EB}
E_{\rm B}(t)= \sum_{M=0}^\infty p_M \sum_{m=0}^M p_{m|M}(t)~Nm \omega_0~,
\end{equation} 
where we introduced the probabilities $p_M \equiv \rho_{M,M}$ and $p_{m|M}(t) \equiv |c_{m,M}(t)|^2$. We fix the initial energy to be $E_{\rm A}=N\omega_0$, which implies the condition $\sum_M M p_M=1$. Introducing also $E_{\rm B}|_N(t) \equiv\sum_{n=0}^M p_{m|M}(t)~Nm \omega_0$, we notice that this quantity cannot exceed $NM\omega_0$,  i.e. $E_{\rm B}|_N(t)\leq NM \omega_0$. Using the condition  $\sum_M M p_M=1$, we obtain $
E_{\rm B}(t)\leq N \omega_0$, which expresses the conservation of the local energy.  We cannot transfer to the battery more than the energy hosted by the charger at the initial time. The energy gained by the battery coincides with this initial energy only in the case of the Fock initial state for which $p_M=\delta_{M,1}$.

\noindent {\color{blue}{\it Classical analog.}}---We now proceed to show that, while in the linear $n=1$ case the quantum dynamics is well described by the classical limit, in the non-linear case the situation is dramatically different. In particular, as we will show momentarily, the classical dynamics of the non-linear model is totally trivial. Once again, this is another strong symptom of a GQA~\cite{QvsC}. Following Ref.~\cite{QvsC}, we therefore compare the quantum model in Eq.~\eqref{H_NL} with the corresponding classical one. In classical Hamiltonian mechanics, the time evolution of a system is defined by Hamilton's equations,  $\Dot{q}_\alpha = \partial_{p_\alpha} \mathcal{H^{\rm cl}} (\boldsymbol{x})$ and $ \Dot{p}_\alpha = -\partial_{q_\alpha} \mathcal{H^{\rm cl}} (\boldsymbol{x})$ where $\boldsymbol{x}$ represents the canonical coordinates $\boldsymbol{x^T} = (\boldsymbol{p}, \boldsymbol{q})$, $\mathcal{H^{\rm cl}}$ is the classical analog Hamiltonian, and $q_\alpha, p_\alpha$ are canonically conjugate variables obeying the Poisson brackets $\{q_\alpha, p_\beta\} = \delta_{\alpha \beta}$ (in contrast to the quantum canonically-conjugate variables $\hat{q}_\alpha, \hat{p}_\alpha$, which fulfil the commutation relation $[\hat q_\alpha, \hat p_\beta] = i \delta_{\alpha \beta}$).

In order to compare quantum and classical systems, we ``reverse'' the canonical quantization procedure. Replacing quantum operators with classical variables we find the classical analog of the quantum model~\eqref{H_NL}:
\begin{align}\label{H_NL|_{cl}}
&\mathcal{H}_{\rm A}^{\rm cl}  = \frac{n \omega_0}{2}\bigg( p_{\rm A}^2 + q_{\rm A}^2 \bigg) ~,\nonumber\\
& \mathcal{H}_{\rm B}^{\rm cl} = \frac{\omega_0}{2} \bigg( p_{\rm B}^2 + q_{\rm B}^2 \bigg) ~,\nonumber\\ 
& \mathcal{H}^{( {\rm cl })}_{\rm n} = \frac{g_n}{2}\left[(q_{\rm A} -ip_{\rm A})(q_{\rm B} + ip_{\rm B})^n + \right. \nonumber \\
& \left.(q_{\rm A} +ip_{\rm A})(q_{\rm B} - ip_{\rm B})^n \right]~,
\end{align}
where $p_{\rm A}$ ($p_{\rm B}$) and $q_{\rm A}$ ($q_{\rm B}$) are the charger (battery) classical conjugate variables.

Consequently, we can derive Hamilton's equations of motion from Eq.~\eqref{H_NL|_{cl}} using the coordinates $X_{\rm A} = q_{\rm A} + i p_{\rm A} $ and $X_B = q_{\rm B} + i p_{\rm B} $. We find $\dot{X}_{\rm A}=-i n \omega_0  X_{\rm A} -i g_n  X_{\rm B}^n$ and 
$\dot{X}_{\rm B}=-i  \omega_0  X_{\rm B} -i n g_n  X_{\rm A} (  X_{\rm B}^*)^{n-1}$.
We emphasize that these equations of motion are not semi-classical approximations to the exact quantum dynamics, but are clearly {\it exact} classical equations of motion for the non-linear battery model described by the {\it classical} Hamiltonian (\ref{H_NL|_{cl}}). 

Using the transformations $\tilde{X}_{\rm A}=e^{in\omega_0t }X_{\rm A}$ and  $\tilde{X}_{\rm B}=e^{i\omega_0t }X_{\rm B}$, we find the classical equations of motion in the rotating reference frame: $\dot{\tilde{X}}_{\rm A}= -ig_n\tilde{X}_{\rm B}^n$ and $\dot{\tilde{X}}_{\rm B}=-in g_n\tilde{X}_{\rm A}(\tilde{X}_{\rm B}^*)^{n-1}$. For $n = 1$, these reduce to the classical equations of motion discussed in Ref.~\cite{QvsC} for the linear model. In this case, the classical battery gains at the end of the charging process exactly the same energy as in the  quantum case (\ref{H_NL}), i.e.~$E^{( 1)}_{\rm B}(\tau) |_{\rm cl} = N\omega_0 \sin^2(g_1\tau)$, which coincides with Eq.~(\ref{stored energyoo}).

On the contrary, it is easy to check by direct inspection that $\tilde{X}_{\rm A}(t)=1$ and $\tilde{X}_{\rm B}(t)=0$ are solutions of the classica equations of motion for $n>1$. Thus:
\begin{equation}
E_{\rm B}(\tau) |_{\rm cl} = 0~.   
\end{equation}
While in the quantum $n>1$ non-linear case the battery charges up (in fact maximally, in the case of an initial Fock state of the charger), in the classical $n>1$ non-linear case no energy is exchanged between the charger and the battery and the latter remains uncharged. Indeed, quantum mechanical fluctuations initiate the charging process, while, classically, the mean value of the coupling term  $\sim \tilde{X}_{\rm B}^{n-1}$ is zero and the dynamics of two systems is effectively decoupled. We therefore have a second important symptom of a GQA: the dynamics in the quantum non-linear case is strikingly different from the dynamics of its classical analog.

\noindent {\color{blue}{\it Certification of the GQA via bounds on power.}}---We finally {\it certificate} the GQA of our non-linear bosonic battery model by using the bounds obtained in Ref.~\cite{JuliaFarre_PRR_2020}. The battery power, indeed, is bounded from above as following~\cite{JuliaFarre_PRR_2020,footnote}:
\begin{equation}
\begin{split}
    \label{boundonpower}
&P_{\rm B}(\tau)\leq P^{\rm bound}_{\rm B}({\tau}) ~,\\
&P^{\rm bound}_{\rm B}({\tau})\equiv 2 \sqrt{\Delta_{\tau} \mathcal{H}_{\rm B}^2 \Delta_{\tau} \mathcal{H}_{\rm NL}^2}~,
\end{split}
\end{equation}
where 
\begin{equation}
\Delta_{\tau} \mathcal{H}_{\rm B}^2  \equiv \frac{1}{\tau} \int_0^\tau dt \langle \delta {\cal H}_{\rm B} \rangle^2_{\psi(t)}~,
\end{equation}
measures the distance traveled in the Hilbert space, with $\ket{\psi(t)}$ being the evolved state in Eq.~\eqref{evolvedstate}, while 
\begin{equation}
\Delta_{\tau} \mathcal{H}_{\rm NL}^2  \equiv \frac{1}{\tau}\int_0^{\tau}dt \langle \delta {\cal H}_{\rm NL} \rangle^2_{\psi(t)}~,
\end{equation}
represents the charging speed of the evolution. Since $\cal{H}_{\rm NL}$ is the total Hamiltonian, its variance is time-independent, i.e.~$\Delta_{\tau} \mathcal{H}_{\rm NL}^2 = \langle \delta {\cal H}_{\rm NL} \rangle^2_{\psi(0)}$. The bound in Eq.~\eqref{boundonpower}---assuming it is tight---allows us to trace the origin of the super-linear scaling mentioned above. Given that $\langle \delta {\cal H}_{\rm NL} \rangle^2_{\psi(0)}$ scales linearly, this variance is not the source of the super-linear scaling observed in Eq.~\eqref{power}. Instead, $\Delta_{\tau} \mathcal{H}_{\rm B}^2$ can enhance the battery performance by allowing entangled states to reduce the trajectory length between $|\psi (0)\rangle_{\rm AB}$ and $|\psi (\tau)\rangle_{\rm AB}$.

Evaluating the quantities in Eq.~\eqref{boundonpower} at the optimal time $\bar{\tau}$, we obtain $\Delta_{\bar{\tau}} \mathcal{H}_{\rm B}^2 =  \omega_0^2 N^2/8$ and $\Delta_{\bar{\tau}} \mathcal{H}_{\rm NL}^2 = g_1^2 N$. We therefore find that 
\begin{equation}\label{eq:bound_on_Power}
 P^{\rm bound}_{\rm B}(\bar{\tau})= \frac{1}{\sqrt{2}}\omega_0 g_1 N^{\frac{3}{2}}~.
\end{equation}
This bound displays a super-linear scaling, with a significant contribution from entanglement, represented by $\Delta_{\bar{\tau}} \mathcal{H}_N^2 \sim N$. To certify GQA using this bound, we consider two scenarios:
i) If the bound in Eq.~\eqref{boundonpower} predicts super-linear scaling due to entanglement generation but the battery power scales only linearly, then the bound is not tight enough to distinguish different contributions.
ii) If the power scales as the bound, then the bound can be used to certify GQA.
In our case, since $P_{\rm B}(\bar{\tau})$ calculated in Eq.\eqref{power} scales as $N^{\frac{3}{2}}$, we are in scenario ii). Therefore, we can use the bound in Eq.\eqref{boundonpower} to certify GQA.
In summary, we have demonstrated in three alternative ways (i.e.~QSLs, classical vs quantum dynamics, and analytical bounds on power) that the quantum battery model described by Eq.~(\ref{H_NL}) displays a GQA. To the best of our knowledge, this is the only quantum battery model where a GQA can be proven analytically. In the case of the SYK batteries, a GQA could only be deduced from essentially-exact numerical studies~\cite{Rossini_PRL_2020} for finite $N$, up to $N=16$.

\noindent  {\color{blue}{\it Experimental implementation}.}---We conclude this work by briefly discussing a potential laboratory platform where the non-linear bosonic model~\eqref{H_NL} can be in principle realized. Following Refs.~\cite{Experimental1, Experimental2, Experimental3}, indeed, we consider two superconducting LC resonators (with inductances $L_1$ and $L_2$ and capacitances $C_1$ and $C_2$) coupled by a Josephson junction~\cite{Experimental1, Experimental2, Experimental3}. The Hamiltonian of this system reads as following
\begin{eqnarray}\label{H_JJ}
\mathcal{H}&=& \omega_1 a^\dagger a+\omega_2 b^\dagger b - E_{\rm J } \cos(\phi_1+\phi_2)~, 
\end{eqnarray} 
where $\omega_i=1/\sqrt{L_iC_i}$, $\phi_1=\lambda_1(a+a^\dagger)$, and $\phi_2=\lambda_2(b+b^\dagger)$. Here, $E_{\rm J }$ is the Josephson energy---which, as explained in Ref.~\cite{Experimental1}, can be tuned by using an external flux $\Phi$---and $\lambda^2_i \equiv 2e^2 \sqrt{L_i/C_i}$ where $e$ is the elementary electron charge. Eq.~(\ref{H_JJ}) corresponds to the zero-bias limit of the one reported by the authors of Ref.~\cite{Experimental1}. Assuming $\lambda_i \ll 1$, we can Taylor-expand the cosine in Eq.~(\ref{H_JJ}) as:
\begin{align}\label{cos}
\cos(\phi_1+\phi_2)=\sum_{m=0}^\infty \sum_{k=0}^m\frac{(-)^m }{k!(2m-k)!} (\phi_1)^k (\phi_2)^{(2m-k)}~.
\end{align} 
Choosing $\omega_2=\omega_0$ and $\omega_1=n\omega_0$ (with $n$ an odd integer) and discarding non-resonant terms in the interaction, we get the following interaction Hamiltonian,
\begin{eqnarray}\label{H_JJ1}
\mathcal{H}_1=E_{\rm J } \frac{(-)^{\frac{n-1}{2}}  }{n!} \lambda_1 \lambda_2^{n}  \big[a^\dagger b^n+a (b^\dagger)^n\big]~,
\end{eqnarray}
which is exactly of the form given in Eq.~\eqref{H_NL}. Note that, in the case $n=1$, Eq.~(\ref{H_JJ1}) reduces to the linear model proposed in Ref.~\cite{QvsC}. Notice that, as Eq.~\eqref{cos} is a perturbative expansion and resonant term are on the order of $\lambda_1 \lambda_2^{n}$, this strategy to realize our non-linear battery model is expected to work only for moderate values of $n$. Indeed, non-resonant terms of the lowest order in  $\lambda_1,\lambda_2$ could become more relevant than  resonant ones in Eq.~\eqref{H_JJ1} if $n\gg 1$. 
In the context of circuit QED, the Hamiltonian in Eq.~\eqref{H_JJ1} is usually associated with the {\it down conversion} phenomenon~ \cite{chang2020,mehta2022}.

In the future, it will be interesting to study the ergotropy~\cite{workext,workext1,workext2,dickebattery} of the model in Eq.~(\ref{H_NL}) and propose a protocol to extract work from the experimental system discussed above.

\noindent  {\color{blue}{\it Acknowledgments}.}---V.S.  acknowledges support from the the National Centre on HPC, Big Data and Quantum Computing - SPOKE 10 (Quantum Computing) and received funding from the European Union Next-GenerationEU - National Recovery and Resilience Plan (NRRP) – MISSION 4 COMPONENT 2, INVESTMENT N. 1.4 – CUP N. I53C22000690001. M.P. was supported by the PNRR MUR project PE0000023-NQSTI (Italy). G.M.A acknowledges useful discussion with S. Gasparinetti, P.A. Erdman and F. Campaioli.

\end{document}